\documentclass[12pt]{article}
\usepackage{amsmath}
\usepackage{amssymb}
\newsymbol \blackbox 1004
\newcommand{\eh}{\hfill}\newlength{\sperr}

\newenvironment{proof}{{\settowidth{\sperr}{\bf\rm Proof}%
\par\addvspace{0.3cm}\noindent\parbox[t]{1.3\sperr}
{\bf\rm P\eh r\eh o\eh o\eh f\eh }%
}}{\nopagebreak\mbox{} $\blackbox$\par\addvspace{0.3cm}}

\def\b{\beta}
\def\g{\gamma}

\def\s{\sigma}
\def\l{\lambda}

\def\O{\Omega}
\def\d{\partial}

\def\vp{\varphi}

\def\wh{\widehat}
\def\wt{\widetilde}
\def\ov{\overline}

\def\BC{{\mathbb C}}
\def\BR{{\mathbb R}}

\newtheorem{Pa}{Paper}[section]
\newtheorem{Tm}[Pa]{{\bf Theorem}}

\newtheorem{Rk}[Pa]{{\bf Remark}}

\newtheorem{Dn}[Pa]{{\bf Definition}}
\newtheorem{Pn}[Pa]{{\bf Proposition}}
\newcommand{\CC}
{{\mathchoice {\setbox0=\hbox{$\displaystyle\rm C$}\hbox{\hbox
to0pt{\kern0.4\wd0\vrule height0.9\ht0\hss}\box0}}
{\setbox0=\hbox{$\textstyle\rm C$}\hbox{\hbox
to0pt{\kern0.4\wd0\vrule height0.9\ht0\hss}\box0}}
{\setbox0=\hbox{$\scriptstyle\rm C$}\hbox{\hbox
to0pt{\kern0.4\wd0\vrule height0.9\ht0\hss}\box0}}
{\setbox0=\hbox{$\scriptscriptstyle\rm C$}\hbox{\hbox
to0pt{\kern0.4\wd0\vrule height0.9\ht0\hss}\box0}}}}

\title{Second harmonic generation: Goursat problem on
the semi-strip and explicit solutions}

\author{Alexander Sakhnovich}

\date{}
\begin{document}
\maketitle

{\bf Short title.} Second harmonic generation \vspace{5mm}

Branch of Hydroacoustics, Marine Institute of Hydrophysics,  \\
National Academy of Sciences of Ukraine \\ e-mail address:
al$_-$sakhnov@yahoo.com

\begin{abstract}
A rigorous and complete solution of the initial-boundary-value
(Goursat) problem for second harmonic generation (and its matrix
analog) on the semi-strip is given in terms of the Weyl functions.
A wide class of the explicit solutions and their Weyl functions is
obtained also.
\end{abstract}

\section{Introduction} \label{intro}
\setcounter{equation}{0}

The second harmonic generation  \cite{G} is one of the simplest
nonlinear interactions and can be presented in the form
\begin{equation} \label{1.1}
\frac{\d}{\d x}u_1=-2 \ov{u_1} u_2, \quad \frac{\d}{\d
t}u_2=u_1^2,
\end{equation}
where $\ov{u_1}$ is complex conjugate to $u_1$. Second harmonic
generation (SHG) (\ref{1.1}) is essential in the study of impulse
propagation. The  results related to the case of a purely
amplitude-modulated fundamental wave go back to Liouville \cite{L,
BS, SMKP}. The SHG integrability was proved and Lax pair was
constructed in \cite{K}. In spite of many important results (see
\cite{AVC, KS1, KhS, SMKP} and references therein) SHG has
remained unsolved. One the reasons for this situation is connected
with the continuously interacting nature of SHG  \cite{KS2}. The
case of the Goursat problem for small and intermediate values of
$x$ and $t$ have been treated in \cite{KS2}, and it is this paper
that attracted our attention to the SHG. It is well-known that
initial-boundary value problems for the integrable nonlinear
equations are both important and   difficult. Several interesting
approaches have been suggested and various results have been
obtained (see, for instance, \cite{B, FI, Kv, KN, KS2, Kis, Kr,
 Sab, Skl}). Here we give a complete solution of the
initial-boundary-value (Goursat) problem for SHG (and its matrix
analog) on the semi-strip using approach that was developed in
\cite{SaL1, SaL2, SaL3} (see also \cite{SaAm}). This approach
consists from two stages: description of the evolution of the Weyl
function in terms of the M\"obius transformation and solution of
the inverse problem - reconstruction of the potential by the Weyl
function.

To study explicit solutions we apply to SHG a version of the
binary iterated B\"acklund-Darboux transformation called GBDT in
the terminology of \cite{SaA3}. B\"acklund transformation is a
well-known and fruitful approach in the nonlinear equations and
spectral theory (see, for instance, \cite{AS, M, MS, Mi, ZM}).
B\"acklund transformations and construction of some self-similar
solutions for SHG have been studied in \cite{KhS, SM, Zh}. GBDT
have been initially developed in \cite{SaA1, SaA2} (see \cite{GKS,
SaA3, SaA5, SaA6} and references therein for various applications)
and provides algebraic representation of the Darboux matrix in the
form of the transform matrix from the system theory, which proves
useful in the study of the explicit SHG solutions.

Some preliminary results and definitions that make the paper
self-contained to a certain degree are given in Section 2. Section
3 is dedicated to the solution of the Goursat problem on the
semi-strip. GBDT for SHG is constructed in Section 4. A wide class
of the explicit solutions and their Weyl functions is obtained in
Section 5.

We denote by $\BR$, $\BC$, and $\BC_+$ the real axis, complex
plane, and open upper half-plane, respectively. Matrix $\b^*$ is
adjoint to $\b$, matrix diag $\{a_1, \, a_2, \ldots \} $ is a
diagonal matrix with the entries $a_1, \, a_2, \ldots$ on the main
diagonal,  $\s$ means spectrum, and
 $I_m$ denotes $m
\times m$ identity matrix ($I_{\cal G}$ is identity operator in
the Hilbert space ${\cal G}$).

\section{Preliminaries} \label{Pr}
\setcounter{equation}{0}

{\bf 2.1.} We shall use the zero curvature representation of our
nonlinear equation:
\begin{equation} \label{2.1}
G_t(x,t,z)-F_x(x,t,z)+[G(x,t,z),F(x,t,z)]=0,
\end{equation}
where $G_t=\frac{\d}{\d t}G$,  $\, [G,F]=G F-F G$ (see \cite{FT}
on this representation, its connections with Lax pairs,
references, and historical remarks). Equation (\ref{2.1}) is the
compatibility condition of the auxiliary systems $w_x=Gw$ and
$w_t=Fw$. Consider the case of the auxiliary systems
\begin{equation} \label{2.2}
\frac{\d}{\d x}w(x,t,z)=G(x,t,z)w(x,t,z), \quad G(x,t,z)=i\big( z
j +j V(x,t) \big),
\end{equation}
\begin{equation} \label{2.3}
\frac{\d}{\d t}w(x,t,z)=F(x,t,z)w(x,t,z), \quad F(x,t,z)=
\frac{i}{z}j H(x,t) ,
\end{equation}
where  $j$, $V$, and $H$ are $2m \times 2m$ matrix functions,
\begin{equation} \label{2.4}
j= \left[
\begin{array}{lr} I_m & 0 \\ 0 & -I_m
\end{array}  \right], \quad V= \left[
\begin{array}{lr} 0 & v \\ v^* & 0
\end{array}  \right], \quad  H(x,t) \geq 0.
\end{equation}
Here system (\ref{2.2}) is the well-known so called Dirac type,
AKNS or ZS system, system (\ref{2.3}) is the well-known canonical
system, and compatibility condition (zero curvature) equation
(\ref{2.1}) is equivalent to the system
\begin{equation} \label{2.5}
H_x(x,t)=i(V(x,t) j H(x,t) - H(x,t)j V(x,t)), \quad i v_t(x,t)=2
\big(H(x,t) \big)_{12},
\end{equation}
where  $\big(H \big)_{kl}$ are $m \times m$ blocks of $H$.

The case
\begin{equation} \label{2.6}
m=1, \quad H(x,t)= \b(x,t)^* \b(x,t), \quad \b(x,t)=[\ov{u_1(x,t)}
\quad u_1(x,t)]
\end{equation}
is of special interest. Notice that when $m=1$, rank $H \, \leq
1$, and $H j H \equiv 0$, then representation (\ref{2.6}) follows
automatically. In this case putting $v=-2i u_2$ we see that
auxiliary systems (\ref{2.2}) and (\ref{2.3}) coincide with the
auxiliary systems in \cite{K}. If $u_1$ is continuously
differentiable in $x$, then  system (\ref{2.5}) is equivalent to
SHG (\ref{1.1}). Indeed, the equivalence of the second equations
in (\ref{1.1}) and (\ref{2.5}) is immediate. Consider the first
equation in (\ref{2.5}). As $H= \b^*\b$ and $\b j \b^*=0$, so the
equality $H_x j \b^*= \b^* \b_x j \b^*$ follows. Therefore
according to (\ref{2.5})  we have $\b^* \b_x j \b^*=-i \b^* \b j V
j \b^*$, i.e., $\b_x j \b^*=-i \b j V j \b^*$. Hence, using once
more $\b j \b^*=0$ (and supposing $\b \not= 0$), we obtain $\b_x
=-i \b j V +f \b$. Now, taking into account the last relation in
(\ref{2.6}) we see that $f= \ov{f}$, and thus $H_x= (\b^*\b)_x=i(V
j H - H j V)+2fH$. Compare this equality with (\ref{2.5}) to
derive $f=0$. In other words we have $\b_x =-i \b j V$, which is
equivalent to $(u_1)_x=-2 \ov{u_1}u_2$, and vice versa $\b_x =-i
\b j V$ yields the first equation in (\ref{2.5}).

{\bf 2.2.} Consider now the $2m \times 2m$ fundamental solution of
the Dirac type auxiliary system with a fixed $t$:
\begin{equation} \label{2.8}
\frac{d}{d x}W(x,z)=i\big( z j +j V(x) \big)W(x,z), \quad
W(0,z)=I_{2m},
\end{equation}
where $ 0 \leq x < \infty$ and $V$ is locally summable.
\begin{Dn} \label{Dn2.1}
A holomorphic function $\varphi$ such that
\begin{equation} \label{2.8'}
\int_0^\infty \left[ \begin{array}{lr} I_m &  i \varphi (z)^*
\end{array} \right]
  K W(x, z)^*
 W(x, z)K^*
 \left[ \begin{array}{c}
I_m \\ - i \varphi (z) \end{array} \right] dx < \infty ,
\end{equation}
where $z \in {\BC}_+$ and
\begin{equation} \label{2.9}
K:=   \frac{1}{\sqrt{2}}       \left[
\begin{array}{cc} I_m &
-I_{m} \\ I_{m} & I_m
\end{array}
\right],    \qquad K^*=K^{-1},
\end{equation}
is called a Weyl function   of system  (\ref{2.8}) on $[0, \,
\infty)$.
\end{Dn}
Weyl functions called also Weyl-Titchmarsh or $m$-functions are
widely used in the spectral theory. For the case of system
(\ref{2.8}) Weyl function admits Herglotz representation
\[
\varphi ( z )= \mu z + \nu + \int_{- \infty }^{ \infty} (
\frac{1}{s- z } - \frac{s}{1+s^{2}})d \tau (s),
\]
where $\mu \geq 0$, $\nu= \nu^*$ and $\tau$ is a nondecreasing $m
\times m$ distribution matrix function. Matrix function $\tau$
proves the  spectral function of the selfadjoint operator
${\mathcal Q}f:= \big(-i j \frac{d}{d x}-V(x) \big)f$ defined on
the proper domain. We shall need some results on system
(\ref{2.8}) and its Weyl functions from \cite{SaA4} (see also
\cite{GKS} for rational Weyl functions and corresponding explicit
formulas). Put
 \begin{equation}       \label{2.10}
{\cal{W}}( z)= {\cal W}(l, z ):=K W(l, \overline{ z})^{*}, \qquad
 {\cal W}( z )=: \{ {\cal W}_{k p}( z )  \}_{k,p=1}^2,
\end{equation}
where ${\cal W}_{kp}$ are the $m \times m$ blocks of $ {\cal W}$.
Denote by ${\cal{N}}(V,l)$ or simply by ${\cal{N}}(l)$ the set of
M\"obius (linear fractional) transformations
 \begin{equation}       \label{2.11}
\varphi (z, l, {\cal P})=i  ({\cal W}_{11}( z ){\cal P}_{1}( z )
+{\cal W}_{12}( z ){\cal P}_{2}( z ))({\cal W}_{21}( z ){\cal
P}_{1}( z ) +{\cal W}_{22}( z ){\cal P}_{2}( z ))^{-1} ,
\end{equation}
where  the pairs ${\cal P}_{1}$ and ${\cal P}_{2}$ are taken from
the often used parameter set of non-singular pairs of meromorphic
$m \times m$ matrix functions with property-$j$, that is the pairs
with the properties
 \begin{equation}       \label{2.12}
{\cal P}_1(z)^*{\cal P}_1(z)+ {\cal P}_2(z)^*{\cal P}_2(z)>0,
\quad {\cal P}_1(z)^*{\cal P}_1(z)- {\cal P}_2(z)^*{\cal P}_2(z)
\geq 0.
\end{equation}
Taking into account formula (1.4) in  \cite{SaA4} one can see that
up to a scalar constant ${\cal{N}}(l)$ here coincides with
${\cal{N}}(l)$ in \cite{SaA4}. From (\ref{2.12}) it follows that
$\det ({\cal W}_{21}( z ){\cal P}_{1}( z ) +{\cal W}_{22}( z
){\cal P}_{2}( z )) \not= 0$ and that
 \begin{equation}       \label{2.12'}
i(\vp(z,l, {\cal P})^*-\vp(z, l, {\cal P}))>0 \quad (z \in \BC_+).
\end{equation}
(see inequalities (2.18), (2.19) \cite{SaA4}). Moreover  the
unique Weyl function $\vp (z)$ exists and is given by the
relations
 \begin{equation}       \label{2.13}
\displaystyle{ \varphi (z)  = \bigcap_{l< \infty} {\cal N}(l)=
\lim_{l_k \to \infty} \vp(z, l_k) }
\end{equation}
for any sequence $\vp_k(z)= \vp (z,l_k) \in {\cal N}(l_k)$. (The
first equality in (\ref{2.13}) follows from Proposition 5.2
\cite{SaA4} and the second follows from the proof of Theorem 2.7
\cite{SaA4}.) Finally when $V$ is locally bounded matrix function,
then Theorem 5.4 and Remark 5.6 in \cite{SaA4} give a well-defined
procedure to recover $V$ by $\vp$ that we shall adduce below. It
is easy to see that $(\zeta +i \eta)^{-2} \varphi (\zeta +i \eta)
\in L^2_{ m \times m}(- \infty , \, \infty)$. Introduce a family
of bounded and boundedly invertible operators $S(l)>0$ ($ 0<l<
\infty$), where $S(l)$  of the form
\begin{equation}       \label{2.14}
S(l)= \frac{d}{d x} \int_0^l s(x-t) \, \cdot \, d t, \quad
s(x)=-s(-x)^*
\end{equation}
acts on $L^2_m(0,l)$, and   the kernel matrix function $s(x)$ is
defined via the Fourier transform
\begin{equation} \label{2.15}
s(x)= \left( \frac{d}{d x} \frac{i}{2 \pi} e^{ \eta x}
 \int_{- \infty}^{\infty}e^{-i \zeta x}
(\zeta +i \eta)^{-2} \varphi (\zeta +i \eta)d \zeta \right)^*,
\quad \eta>0.
\end{equation}
Introduce further $m \times 2m$ matrix function $\g$ by the
relation
\begin{equation} \label{2.16}
\g(x)= \frac{1}{\sqrt{2}} \left( [-I_m \quad I_m] - \int_0^x
(s^{\prime}(t))^* S(x)^{-1}[  s(t) \quad I_m]d t \right),
\end{equation}
where operator $S^{-1}$ is acting on $[  s(t) \quad I_m]$
columnwise. Now potential $v$ (and thus $V$) is recovered by the
equality
\begin{equation}\label{2.17}
v(x)=2i \chi_{x}(2x) J \g(2x)^*  \quad \left( J=\left[
\begin{array}{lr}0 & I_m  \\ I_m & 0
\end{array}  \right] \right).
\end{equation}
Here $m \times 2m$ matrix function $\chi$ is uniquely and easily
obtained via $\g$ by the properties
\begin{equation}\label{2.18}
 \chi(0)=\frac{1}{ \sqrt{2}}[I_m \quad
I_m], \quad \chi_x(x) J \chi(x)^* \equiv 0, \quad \chi(x) J
\g(x)^* \equiv 0.
\end{equation}
One can recover the potential of system (\ref{2.8}) in a more
traditional way - via spectral function, but the direct recovery
via Weyl function is a more general method that is applicable also
in the case of the skew-self-adjoint analog of (\ref{2.8}) as in
\cite{SaAm}. Procedure (\ref{2.14})-(\ref{2.18}) is closely
related to the study of the high energy asymptotics of the Weyl
functions in \cite{CG, GS, S} as well.

{\bf 2.3.} Suppose now that matrix functions $G(x,t,z)$ and
$F(x,t,z)$ given in (\ref{2.2}) and (\ref{2.3}) are continuously
differentiable in the domain $0 \leq x<l$, $\quad 0 \leq t<T$
($-T<t \leq 0$) and that (\ref{2.1}) holds. Similar to (\ref{2.8})
we shall use notation $W(x,t,z)$ for the fundamental solution of
(\ref{2.2}) normalized by the condition $W(0,t,z)=I_{2m}$ and we
shall denote by $R(x,t,z)$ the fundamental solution of
(\ref{2.3}):
\begin{equation}       \label{2.19}
\frac{\d}{\d t}R(x,t,z)=\frac{i}{z}j H(x,t)R(x,t,z), \quad
 R(x,0,z)=I_{2m}.
\end{equation}
It easily follows (see  \cite{SaL3}, p.168) that
\begin{equation}       \label{2.20}
W(x,t,z)= R(x,t,z)W(x,0,z)R(0,t,z)^{-1}.
\end{equation}
\section{Goursat problem} \label{GP}
\setcounter{equation}{0}

Denote now by $\vp(t, z)$ the Weyl function of system (\ref{2.8})
with $V(x)=V(x,t)$ and put
\begin{equation}       \label{3.1}
{\cal{R}}(t, z)= \{ {\cal R}_{k p}(t, z )  \}_{k,p=1}^2:=K \big(
R(0,t, \overline{ z})^{*} \big)^{-1}K^*,
\end{equation}
where ${\cal R}_{k p}$ are $m \times m$ blocks of ${\cal R}$.
\begin{Tm} \label{Tm3.1}
Suppose $2m \times 2m$ matrix function $H(x,t) \geq 0$
and $m \times m$ matrix function $v(x,t)$ are continuously
differentiable in the semi-strip $D= \{ (x,t): \,  0 \leq x<
\infty , \, 0 \leq t< T \}$ and satisfy equations (\ref{2.5}) with
$V$ given by the second relation in (\ref{2.4}). Then the
evolution $\vp(t, z)$ of the Weyl function is given by the formula
\begin{equation}       \label{3.2}
\vp(t, z)=i\big( -i {\cal R}_{11}(t, z )\vp(0, z)+ {\cal R}_{1
2}(t, z ) \big) \big( -i {\cal R}_{21}(t, z ) \vp(0, z) + {\cal
R}_{22}(t, z ) \big)^{-1}.
\end{equation}
Moreover $H$ and $v$ in the semi-strip $D$ are uniquely recovered
by the initial-boundary values $v(x,0)$ and $H(0,t)$.  Here
$\vp(0,z)$ is defined via $v(x,0)$ by the formula (\ref{2.13}),
${\cal R}(t,z)$ is defined via $H(0,t)$ by the formulas
(\ref{2.19}) and (\ref{3.1}), evolution $\vp(t,z)$ of the Weyl
function follows now from (\ref{3.2}) and, finally, potential
$v(x,t)$ is obtained via $\vp(t,z)$ by the procedure
(\ref{2.14})-(\ref{2.18}). After $v$ is recovered we get $H(x,t)$
as a unique solution with the initial value $H(0,t)$ of the linear
system $H_x=i(V j H - H j V)$.
\end{Tm}
\begin{proof}. In view of the results cited in Section \ref{Pr}
(subsection 2.2) we need to prove only evolution formula
(\ref{3.2}). For that purpose  we  slightly modify  the proof of
Theorem 1.1 in Ch. 12 \cite{SaL3}. Rewrite (\ref{2.20}) in the
form
\begin{equation}       \label{3.3}
R(l,t,z)^{-1}W(l,t,z)K^*=W(l,0,z)K^* K R(0,t,z)^{-1}K^*
\end{equation}
Replace $z$ by $\ov{z}$, take adjoints of both sides in
(\ref{3.3}) and use  definitions (\ref{2.10}) and (\ref{3.1}) to
get
\begin{equation}       \label{3.4}
{\cal U}(l,t,z)= \{ {\cal U}_{kp}(l,t,z) \}_{k,p=1}^2:= {\cal
W}(l,t,z)\big( R(l,t,\ov{z})^{-1} \big)^*= {\cal R}(t,z){\cal
W}(l,0,z),
\end{equation}
where ${\cal U}_{kp}$ are $m \times m$ blocks.  Choose now a
non-singular pair $\wh {\cal P}_1$, $\wh {\cal P}_2$ with
property-$j$ and  consider a M\"obius transformation
\[ \psi(l,t,z):=({\cal U}_{11}(l,t,z) \wh {\cal P}_1(z) +
{\cal U}_{12}(l,t,z)\wh {\cal P}_2(z)) \times
\]
\begin{equation}       \label{3.4'}
( {\cal U}_{21}(l,t,z) \wh {\cal P}_1(z) + {\cal U}_{22}(l,t,z)
\wh {\cal P}_2(z))^{-1}.
\end{equation}
Notice that by (\ref{2.19}) for $z \in \BC_+$ we have
\begin{equation}       \label{3.5}
\frac{\d }{\d t} \big(R(x,t, \ov{z})^*j R(x,t, \ov{z}) \big)=
\frac{i(z- \ov{z}) }{|z|^2}R(x,t, \ov{z})^*H(x,t) R(x,t, \ov{z})
\leq 0.
\end{equation}
Taking into account that $R(x,0, \ov{z})^*j R(x,0, \ov{z})=j$ we
derive from (\ref{3.5}) inequality:
\begin{equation}       \label{3.6}
R(l,t, \ov{z})^*j R(l,t, \ov{z}) \leq j \quad (z \in \BC_+).
\end{equation}
From (\ref{3.6}) it follows that
\begin{equation}       \label{3.7}
 R(l,t,\ov{z})^{-1}  j \big( R(l,t,\ov{z})^{-1} \big)^* \geq j
\quad (z \in \BC_+),
\end{equation}
and therefore the pair
\begin{equation}       \label{3.8}
 \left[
\begin{array}{c} {\cal P}_1(z) \\ {\cal P}_2(z)
\end{array}  \right]:= \big( R(l,t,\ov{z})^{-1} \big)^*\left[
\begin{array}{c} \wh {\cal P}_1(z) \\ \wh {\cal P}_2(z)
\end{array}  \right]
\end{equation}
satisfies (\ref{2.12}). According to definitions of ${\cal U}$ and
$\psi$ and formula (\ref{3.8}) we obtain
\[
\psi(l,t,z)= ({\cal W}_{11}(l,t, z ){\cal P}_{1}( z ) +{\cal
W}_{12}(l,t, z ){\cal P}_{2}( z )) \times
\]
\begin{equation}       \label{3.9}
({\cal W}_{21}(l,t, z ){\cal P}_{1}( z ) +{\cal W}_{22}(l,t,  z
){\cal P}_{2}( z ))^{-1}.
\end{equation}
As the pair ${\cal P}_1$, ${\cal P}_2$  satisfies (\ref{2.12}), so
by (\ref{2.13}) and (\ref{3.9}) we get
\begin{equation}       \label{3.10}
 \lim_{l \to \infty} \, \psi(l,t,z) =-i \vp(t,z).
\end{equation}
On the other hand from the last equality in (\ref{3.4}) and
(\ref{3.4'}) we get also
\begin{equation}       \label{3.11}
\psi(l,t,z)= \big( -i {\cal R}_{11}(t, z )  \vp_l( z)+ {\cal R}_{1
2}(t, z ) \big) \big( -i {\cal R}_{21}(t, z )  \vp_l( z) + {\cal
R}_{22}(t, z ) \big)^{-1},
\end{equation}
where $\vp_l \in {\cal N}(l,0)$ is given by the formula
\[ \vp_l( z)=i
({\cal W}_{11}(l,0, z ) \wh {\cal P}_{1}( z ) +{\cal W}_{12}(l,0,
z ) \wh {\cal P}_{2}( z )) \times \]
\begin{equation}       \label{3.12}
({\cal W}_{21}(l,0, z )\wh {\cal P}_{1}( z ) +{\cal W}_{22}(l,0, z
) \wh {\cal P}_{2}( z ))^{-1}.
\end{equation}
By (\ref{2.13}) and (\ref{3.12}) we have
\begin{equation}       \label{3.13}
\lim_{l \to \infty} \vp_l( z)= \vp(0, z).
\end{equation}
Supposing
\begin{equation}       \label{3.14}
\det \, \big( -i {\cal R}_{21}(t, z ) \vp(0, z) + {\cal R}_{22}(t,
z ) \big) \not= 0,
\end{equation}
and taking into account (\ref{3.11}) and (\ref{3.13}) we see that
\[
\lim_{l \to \infty} \psi(l,t,z)= \big( -i {\cal R}_{11}(t, z )
\vp(0, z)+ {\cal R}_{1 2}(t, z ) \big) \times
\]
\begin{equation}       \label{3.15}
 \big( -i {\cal R}_{21}(t, z
) \vp(0, z) + {\cal R}_{22}(t, z ) \big)^{-1}.
\end{equation}
Compare (\ref{3.10}) and (\ref{3.15}) to obtain (\ref{3.2}).

Finally, using definitions in (\ref{2.4}), (\ref{2.9}),
 and (\ref{2.17}) one can easily check that
\begin{equation}       \label{3.16}
K^*J K=j, \quad K j K^*=J.
\end{equation}
Hence in view of  (\ref{3.1}) and (\ref{3.7}) it follows that
${\cal R}(t,z)^*J {\cal R}(t,z) \geq J$, i.e.,
\begin{equation}       \label{3.17}
[i \vp(0,z)^* \quad I_m]{\cal R}(t,z)^*J {\cal R}(t,z)\left[
\begin{array}{c} -i \vp(0,z) \\ I_m
\end{array}  \right] \geq i(\vp(0,z)^*-\vp(0,z)),
\end{equation}
$z \in \BC_+$. Recall that $i(\vp(0,z)^*-\vp(0,z))>0$ and so
(\ref{3.17}) yields (\ref{3.14}).
\end{proof}
In a similar way the Goursat problem on the semi-strip $\wh D= \{
(x,t): \,  0 \leq x< \infty , \, -T< t \leq 0 \}$ can be solved.
\begin{Tm} \label{Tm3.2}
Suppose $2m \times 2m$ matrix function $H(x,t) \geq 0$ and $m
\times m$ matrix function $v(x,t)$ are continuously differentiable
in the semi-strip $\wh D$ and satisfy equations (\ref{2.5}) with
$V$ given by the second relation in (\ref{2.4}). Then $H$ and $v$
in  $\wh D$ are uniquely recovered by the initial-boundary values
$v(x,0)$ and $H(0,t)$.  Here $\vp(0,z)$ is defined via $v(x,0)$ by
the formula (\ref{2.13}), ${\cal R}(t,z)$ is defined via $H(0,t)$
by the formulas (\ref{2.19}) and (\ref{3.1}), Weyl functions
$\vp(t,z)$ are given by (\ref{3.2}) and, finally, potential
$v(x,t)$ is obtained via $\vp(t,z)$ by the procedure
(\ref{2.14})-(\ref{2.18}). After $v$ is recovered we get $H(x,t)$
as a unique solution with the initial value $H(0,t)$ of the linear
system $H_x=i(V j H - H j V)$.
\end{Tm}
\begin{proof}. Notice that instead of (\ref{3.6}) formula
(\ref{3.5}) yields now inequality
\begin{equation}       \label{3.18}
R(l,t, \ov{z})^*j R(l,t, \ov{z}) \geq j \quad (z \in \BC_+, \quad
t \leq 0 ),
\end{equation}
i.e., $R$ and $R^*$ are $j$-expanding. Analogously we have now
\begin{equation}       \label{3.19}
\big({\cal R}(t, z)^{-1}\big)^*J {\cal R}(t, z)^{-1} \geq J \quad
(z \in \BC_+, \quad t \leq 0 ).
\end{equation}
Therefore we modify  formula  (\ref{3.4}):
\begin{equation}       \label{3.20}
{\cal U}(l,t,z)= \{ {\cal U}_{kp}(l,t,z) \}_{k,p=1}^2:={\cal R}(t,
z)^{-1} {\cal W}(l,t,z)= {\cal W}(l,0,z)R(l,t, \ov{z})^*.
\end{equation}
Consider again $\lim_{l \to \infty} \psi(l,t,z)$, where $\psi$ is
given by (\ref{3.4'}), and use (\ref{3.20}) to get
\begin{equation}       \label{3.21}
i\big( -i {\cal T}_{11}(t, z )\vp(t, z)+ {\cal T}_{1 2}(t, z )
\big) \big( -i {\cal T}_{21}(t, z ) \vp(t, z) + {\cal T}_{22}(t, z
) \big)^{-1}=\vp(0, z),
\end{equation}
where ${\cal T}(t,z)= \{ {\cal T}_{kp}(t,z) \}_{k,p=1}^2:={\cal
R}(t, z)^{-1}$. Rewrite (\ref{3.21}) as
\begin{equation}       \label{3.22}
{\cal T}(t,z) \left[
\begin{array}{c} -i \vp(t, z) \\ I_m
\end{array}  \right]= \left[
\begin{array}{c} -i \vp(0, z) \\ I_m
\end{array}  \right] c(z), \quad \det \, c(z) \not=0 \quad ({\cal T}
={\cal R}^{-1}).
\end{equation}
Multiply both sides of (\ref{3.22}) by ${\cal R}$ to obtain
(\ref{3.14}) and, finally, (\ref{3.2}).
\end{proof}

\section{GBDT for second harmonic generation} \label{GBDT}
\setcounter{equation}{0}

A general result on the GBDT for systems rationally depending on
$\l$ (and spectral parameter $\l$ depending on the variables $x$
and $t$) have been proved in \cite{SaAg}. Systems polynomially
depending on $\l$ and $\l^{-1}$ have been in greater detail
treated in \cite{SaA3}. Here we shall need a reduction of Theorem
1.1 \cite{SaA3} for the  $2m \times 2m$ first order system of the
form
\begin{equation} \label{4.1}
\frac{d}{d s} w(s, \lambda)+ (\l q_1(s)+q_0(s)+ \l^{-1}q_{-1}(s))
w(s, \lambda )=0,
\end{equation}
 where the coefficients $ q_{k}(s)$ are $2m \times 2m$ locally summable
 on $[0,$ $ c)$  $(c \leq \infty )$ matrix
functions, and the equalities
\begin{equation} \label{4.2}
q_p(s)^*=-j q_p j \quad (p=1, \, 0 , \, -1)
\end{equation}
hold.  After fixing an integer $n>0$ the GBDT of the system
(\ref{4.1}) is determined by the three parameter matrices: two $n
\times n$ matrices $A$ and $S(0)$, and $n \times 2m$ matrix
 $\Pi(0)$ such that
\begin{equation} \label{4.3}
A S(0)-S(0)A^* =i \Pi(0) j  \Pi(0)^{*}, \quad \det \, A \not=0,
\quad S(0)=S(0)^*.
\end{equation}
Given these parameter matrices we define matrix function $\Pi(s)$
by its initial value $\Pi(0)$ and linear differential system
\begin{equation} \label{4.4}
\frac{d}{d s} \Pi(s)= A \Pi(s) q_1(s)+ \Pi(s) q_0(s)+ A^{-1}
\Pi(s) q_{-1}(s),
\end{equation}
i.e., $\Pi$ is a generalized eigenfunction of the dual to
(\ref{4.1}) system $\frac{d}{d s} w= \l w q_1+ wq_0+ \l^{-1}w
q_{-1}$. Matrix function $S(s)$ is defined by $S(0)$ and
$\frac{d}{d s}S$:
\begin{equation} \label{4.5}
\frac{d}{d s}S=i \Big( \Pi(s)q_1(s)j \Pi(s)^*- A^{-1} \Pi(s)
q_{-1}(s) j \Pi(s)^* \big(A^* \big)^{-1} \Big).
\end{equation}
Notice that relations (\ref{4.3})-(\ref{4.5}) are chosen so that
\begin{equation} \label{4.6}
AS(s)-S(s)A^*=i \Pi(s) j \Pi(s)^{*}, \quad S(s)=S(s)^*.
\end{equation}
We introduce Darboux matrix (gauge transformation) by the formula
\begin{equation} \label{4.7}
w_{A}(s, \lambda )=I_{2m}- i j \Pi(s)^{*}S(s)^{-1}(A- \lambda
I_{n})^{-1} \Pi(s).
\end{equation}
Compare (\ref{4.5}) with subsection 2.2, where the operators $S$
necessary to solve the general type inverse problem have been
defined differently, - there is difference between the usage of
the transfer matrix function and ways to define its elements for
the general type problems and for the case of the explicit
solutions, i.e., the GBDT case. By Theorem 1.1 (see also
Proposition 1.4) \cite{SaA3} we have
\begin{Tm} \label{TmGBDT}
Suppose coefficients of system (\ref{4.1}) satisfy (\ref{4.2}) and
the parameter matrices $A$, $S(0)$ and $\Pi(0)$ satisfy relations
(\ref{4.3}). Define matrix functions  $\Pi$ and $S$ by the
equations (\ref{4.4}) and (\ref{4.5}). Then in the points of
invertibility of $S$ the matrix function $w_A$ satisfies the
system
\[
\frac{d}{d s}w_A(s, \lambda )= w_A(s, \lambda )(\l q_1(s)+q_0(s)+
\l^{-1}q_{-1}(s)) -
\]
\begin{equation} \label{4.8}
(\l \wt q_1(s)+ \wt q_0(s)+ \l^{-1} \wt q_{-1}(s))w_A(s, \l),
\end{equation}
where
\begin{equation} \label{4.9}
\wt q_1 \equiv q_1, \quad \wt q_0(s)= q_0(s)-(q_1(s)
Y_0(s)-X_0(s)q_1(s)),
\end{equation}
\begin{equation} \label{4.9'}
 \wt q_{-1}(s)=q_{-1}(s)+q_{-1}(s)
Y_{-1}(s)-X_{-1}(s)q_{-1}(s)-X_{-1}(s)q_{-1}(s)Y_{-1}(s),
\end{equation}
\begin{equation} \label{4.10}
 X_{k}(s) =i j \Pi(s) ^{*}S(s) ^{-1}A^{k} \Pi(s), \quad
Y_{k}(s) = i j \Pi(s) ^{*} \big(A^* \big)^{k}S(s) ^{-1} \Pi(s).
\end{equation}
Moreover we have
\begin{equation} \label{4.10'}
\wt q_p(s)^*=-j \wt q_p j \quad (p=1, \, 0 , \, -1).
\end{equation}
If $w$ satisfies system (\ref{4.1}), then matrix function
\begin{equation} \label{4.11}
\wt w(s, \l):= w_A(s, \l)w(s, \l)
\end{equation}
satisfies GBD transformed system
\begin{equation} \label{4.12}
\frac{d}{d s} \wt w(s, \lambda)+ (\l \wt q_1(s)+\wt q_0(s)+
\l^{-1} \wt q_{-1}(s))\wt w(s, \lambda )=0.
\end{equation}
\end{Tm}
 Transfer matrix functions of the form
 \[
 w_A(\l)=I_{\cal G}- \Pi_2^*S^{-1}(A_1 - \l I_{\cal H})^{-1} \Pi_1
 \quad (A_1S-SA_2= \Pi_1 \Pi_2^*)
 \]
have been introduced and studied by L. Sakhnovich in the context
of his method of operator identities (see \cite{SaL2, SaL3} and
references therein) and take roots in the M.~S. Liv\v{s}ic
characteristic matrix functions. From (\ref{4.7}) easily follows
\cite{SaL2, SaL3} a useful equality
\begin{equation} \label{4.13}
w_A(s, \ov{\l})^*j w_A(s, \l)=j.
\end{equation}
Notice that according to  (\ref{4.9}) identities $q_1 \equiv 0$
and $q_0 \equiv 0$ yield $ \wt q_1 \equiv 0$ and $\wt q_0 \equiv
0$. By (\ref{4.7}) and (\ref{4.10}) identity (\ref{4.9'}) can be
rewritten as
\begin{equation} \label{4.14}
\wt q_{-1}(s)=(I_{2m}-X_{-1}(s))q_{-1}(s)(I_{2m}+Y_{-1}(s))=
w_A(s, 0)q_{-1}(s)j w_A(s, 0)^*j.
\end{equation}
If $q_{-1} \equiv 0$, then we have $ \wt q_{-1} \equiv 0$ also. It
is immediate that   auxiliary systems (\ref{2.2}) and (\ref{2.3})
satisfy conditions of Theorem \ref{TmGBDT}. To apply Theorem
\ref{TmGBDT} we consider matrix functions $\Pi$, $S$, $w_A$,
$X_k$, and $Y_k$ that depend on the variables $x$ and $t$ instead
of one variable $s$. Namely, we fix $n>0$ and parameter matrices
$A$, $S(0,0)$, and $\Pi(0,0)$ such that
\begin{equation} \label{4.15}
A S(0,0)-S(0,0)A^* =i \Pi(0,0) j  \Pi(0,0)^{*}, \quad \det \, A
\not=0, \quad S(0,0)=S(0,0)^*,
\end{equation}
and introduce matrix functions $\Pi(x,t)$ and $S(x,t)$ by the
equations
\begin{equation} \label{4.16}
 \Pi_x(x,t)=-i A \Pi(x,t)j-i \Pi(x,t) j V(x,t), \quad  \Pi_t(x,t)=-i A^{-1}
\Pi(x,t) j H(x,t),
\end{equation}
\begin{equation} \label{4.17}
S_x(x,t)=  \Pi(x,t) \Pi(x,t)^*, \quad S_t(x,t)=- A^{-1} \Pi(x,t)j
H(x,t) j \Pi(x,t)^* \big(A^* \big)^{-1} .
\end{equation}
Compatibility of systems (\ref{4.16}) and (\ref{4.17}) follows
from (\ref{2.5}) (see \cite{SaA3}). Now Theorem \ref{TmGBDT}
yields the following result.
\begin{Tm} \label{Tm4.2}
Suppose continuously differentiable matrix functions $H$ and $v$
satisfy system (\ref{2.5}). Choose $n>0$ and parameter matrices
$A$, $S(0,0)$, and $\Pi(0,0)$ such that (\ref{4.15}) holds. Then
matrix function
\begin{equation} \label{4.18}
\wt H(x,t):=j w_A(x,t, 0)j H(x,t) j w_A(x,t, 0)^*j,
\end{equation}
where $w_A(x,t, \l)=I_{2m}- i j \Pi(x,t)^{*}S(x,t)^{-1}(A- \lambda
I_{n})^{-1} \Pi(x,t)$, and matrix function
\begin{equation} \label{4.19}
 \wt
v(x,t)=v(x,t)-2 \big( X_0(x,t) \big)_{12}
 \quad( X_{0}(x,t) =i j \Pi(x,t) ^{*}S(x,t)
^{-1} \Pi(x,t))
\end{equation}
satisfy system (\ref{2.5}) also.  Moreover, if
\begin{equation} \label{4.20}
H(x,t) \geq 0, \quad H(x,t)j H(x,t) \equiv 0,
\end{equation}
then we have
\begin{equation} \label{4.21}
\wt H(x,t) \geq 0, \quad \wt H(x,t)j \wt H(x,t) \equiv 0.
\end{equation}
\end{Tm}
\begin{proof}.
As system (\ref{2.5}) is equivalent to the compatibility condition
(\ref{2.1}) so there is a non-degenerate $m \times m$ matrix
function $w$ that satisfies equations $w_x=Gw$, $w_t=Fw$. From the
identities (\ref{4.6}) and (\ref{4.15}) in the case of two
variables we get
\begin{equation} \label{4.21'}
A S(x,t)-S(x,t) A^*=i \Pi(x,t)j \Pi(x,t)^*.
\end{equation}
Substitute now $x=s$ into Theorem \ref{TmGBDT} to get $\wt w_x=(i
z j- \wt q_0) \wt w$ for $\wt w =w_Aw$. Taking into account that
$q_0=-i j V$ and $q_1=-i j$ we rewrite the second equality in
(\ref{4.9}) as $\wt q_0=-i j V+(j X_0-X_0j)$, i.e.,
\[
\wt q_0=-i j \wt V, \quad \wt V= V+j X_0 j- X_0= \left[
\begin{array}{lr} 0 & \wt v \\ \wt v^* & 0
\end{array}  \right].
\]
Therefore we get
\begin{equation} \label{4.22}
\frac{\d}{\d x} \wt w(x,t, \l)=i \big(z j+j \wt V(x,t) \big)\wt
w(x,t, \l), \quad \wt v(x,t)=v(x,t)-2 \big( X_0(x,t) \big)_{12}.
\end{equation}
 Substitute also $t=s$ into Theorem \ref{TmGBDT}
to get $\wt w_t=-z^{-1} \wt q_{-1} \wt w$. Taking into account
that $q_{-1}= -i j H$ from formula (\ref{4.14}) it follows that
$\wt q_{-1}= -i j \wt H$, where $\wt H$ is given by the equality
(\ref{4.18}). In other words we have
\begin{equation} \label{4.23}
\frac{\d}{\d t}  \wt w(x,t, \l)= \frac{i}{z} j \wt H(x,t)  \wt
w(x,t, \l).
\end{equation}
The compatibility  condition for systems (\ref{4.22}) and
(\ref{4.23}) is equivalent to the system
\begin{equation} \label{4.25}
\wt H_x(x,t)=i( \wt V(x,t) j \wt H(x,t) - \wt H(x,t)j \wt V(x,t)),
\quad i \wt v_t(x,t)=2 \big( \wt H(x,t) \big)_{12}.
\end{equation}
Finally relations (\ref{4.21}) follow from (\ref{4.13}),
(\ref{4.18}), and (\ref{4.20}).
\end{proof}
\section{Explicit solutions} \label{ES}
\setcounter{equation}{0}

In this section we shall treat the case $m=1$, $H= \b^* \b$,
\begin{equation} \label{5.1}
 \b(x,t)=[ \ov{b} e^{-i(cx+dt)} \quad b e^{i(cx+dt)}],
\quad v(x,t)=- \frac{b^2}{d}e^{2i(cx+dt)}, \quad c d=|b|^2,
\end{equation}
where $c,d \, \in \BR$. In other words that is the case
\begin{equation} \label{5.2}
u_1(x,t)=b e^{i(cx+dt)}, \quad u_2(x,t)= \frac{b^2}{2 i
d}e^{2i(cx+dt)}.
\end{equation}
It is immediate that these $H$ and $v$ satisfy system (\ref{2.5}),
i.e., $u_1$ and $u_2$ satisfy SHG. To construct $\Pi$
corresponding to the initial solution given by (\ref{5.1}) we need
$n \times n$ matrices $Q_1$ and $Q_2$ such that
\begin{equation} \label{5.3}
Q_2^2=d^2(I_n-2c A^{-1}), \quad AQ_2= Q_2 A, \quad Q_1=-
d^{-1}AQ_2.
\end{equation}
Define columns $\Phi$ and $\Psi$ of $\Pi =[ \Phi \quad \Psi]$ by
the relations
\begin{equation} \label{5.4}
\Phi(x,t)=e^{-i(cx+dt)} \big(e(x,t)f_1 + e(x,t)^{-1}f_2 \big),
\quad e(x,t):= \exp \{ i(xQ_1+tQ_2) \},
\end{equation}
\begin{equation} \label{5.5}
\Psi(x,t)=- d  \ov{b}^{-2}e^{i(cx+dt)} \big(e(x,t)(A-cI_n+Q_1)f_1
+ e(x,t)^{-1}(A-cI_n-Q_1)f_2 \big).
\end{equation}
Direct calculation shows that $\Pi$ satisfies (\ref{4.16}). Now in
view of (\ref{4.17}) we obtain
\begin{equation} \label{5.6}
S(x,t)=S(0,t)+ \int_0^x \Pi(s,t) \Pi(s,t)^*d s,
\end{equation}
\begin{equation} \label{5.7}
S(0,t)=S(0,0)- \int_0^t  A^{-1} \Pi(0,s)j H(0,s) j \Pi(0,s)^*
\big(A^* \big)^{-1} d s.
\end{equation}
\begin{Rk} \label{Rk5.1} If $A$ is diagonal (or similar to
diagonal) matrices $Q_1$ and $Q_2$ always exist and are easy to
construct. Moreover if we  require additionally that $\s(A) \cap
\s(A^*)= \emptyset$, then $S(x,t)$ is uniquely recovered from the
identity (\ref{4.21'}). Namely, for $A=U {\mathrm{diag}} \{ a_1,
\, a_2, \ldots \} U^{-1}$ formula (\ref{4.21'}) yields
\begin{equation} \label{5.8}
S(x,t)=i U \Big\{ (a_k- \ov{a}_p)^{-1} \big( U^{-1} \Pi(x,t)j
\Pi(x,t)^* (U^*)^{-1} \big)_{kp} \Big\}_{k,p=1}^n U^*.
\end{equation}
\end{Rk}
The next proposition on the explicit solutions is a corollary of
Theorem \ref{Tm4.2}.
\begin{Pn} \label{Pn5.2} Let   functions $H= \b^* \b$ and $v$
be given by (\ref{5.1}), matrix functions $\Pi=[ \Phi \quad \Psi]$
be given by (\ref{5.3})-(\ref{5.5}) and $S$ be given by
(\ref{5.6}) and (\ref{5.7}) (or by (\ref{5.8}) if the conditions
of Remark \ref{Rk5.1} hold). Let also relations (\ref{4.15}) be
valid. Then in the points of invertibility of $S$  functions $\wt
H= \wt \b^* \wt \b$ and $\wt v= v-2i \Phi^*S^{-1} \Psi$, where
\[
\wt \b =[\wt \b_1 \quad \wt \b_2]= \b (j+ij \Pi^*(A^*)^{-1}S^{-1}
\Pi) , \] satisfy system (\ref{2.5}).
\end{Pn}
In particular, functions $\wt u_1= \sqrt{\ov{\wt \b_1} \wt \b_2}$
and $\wt u_2=u_2+\Phi^*S^{-1} \Psi$ satisfy SHG in the domains,
where $S$ is invertible and branch $\sqrt{\ov{\wt \b_1} \wt \b_2}$
is continuously differentiable. Notice further that choosing
$S(0,0)>0$ in view of (\ref{5.6}) and (\ref{5.7}) we get
\begin{equation} \label{5.9}
S(x,t)>0 \quad (x \geq 0, \quad t \leq 0),
\end{equation}
i.e., for $x \geq 0, \quad t \leq 0$ matrix function $S$ is
invertible. In this case we obtain Weyl functions $\vp(t,z)$ of
systems
\begin{equation} \label{5.10}
\frac{d}{d x}\wt W(x,t,z)=i\big( z j +j \left[
\begin{array}{lr} 0 & \wt v(x,t) \\ \wt v(x,t)^* & 0
\end{array}  \right] \big)\wt W(x,t,z) \quad (x \geq
0)
\end{equation}
explicitly also.
\begin{Pn} \label{Pn5.3} Suppose conditions of Proposition
\ref{Pn5.2} are satisfied and $S(0,0)>0$. Then Weyl functions of
systems (\ref{5.10})  are given by the equality
\begin{equation} \label{5.11}
\vp(t,z)=i \frac{\O_2(t,z)}{ \O_1(t,z)} \quad (t \leq 0),
\end{equation}
where functions $\O_1$ and $\O_2$ have the form:
\begin{equation} \label{5.12}
\O(t,z)= \left[
\begin{array}{c} \O_1(t,z) \\ \O_2(t,z)
\end{array}  \right]:=K w_A(0,t,z) e^{idtj} Z \left[
\begin{array}{c} 1 \\ 0
\end{array}  \right], \quad Z:= \left[
\begin{array}{lr} 1& 1  \\ g_1(z) & g_2(z)
\end{array}  \right],
\end{equation}
\begin{equation} \label{5.13}
g_1(z):= db^{-2}(z-c-h(z)), \quad g_2(z):= db^{-2}(z-c+h(z)),
\end{equation}
\begin{equation} \label{5.14}
h(z):= \sqrt{z(z-2c)}, \quad (z \in \BC_+, \quad \Im h >0).
\end{equation}
\end{Pn}
\begin{proof}. According to Definition \ref{Dn2.1} it will suffice
to show that
\[
\wt W(x,t,z)K^* \O(t,z) \in L^2_2(0, \, \infty) \]
(is squarely
summable) and $\O_1 \not= 0$. For this purpose notice that by
Theorem \ref{TmGBDT} the normalized fundamental solution $\wt W$
($\wt W(0,t,z)=I_2$) admits representation
\begin{equation} \label{5.15}
\wt W(x,t,z)=w_A(x,t,z)W(x,t,z)w_A(0,t,z)^{-1},
\end{equation}
where $W$ is the normalized solution of the initial system
$W_x=i(z j+j V)W$ with $v$ as in (\ref{5.1}). One can compute
directly that
\begin{equation} \label{5.16}
W(x,t,z)=e^{i(cx+dt)j}Ze^{i h(z) x j} Z^{-1} e^{-idtj}.
\end{equation}
Therefore in view of (\ref{5.12}), (\ref{5.15}), and (\ref{5.16})
we have
\begin{equation} \label{5.17}
\wt W(x,t,z)K^* \O(t,z)=e^{ih(z)x}w_A(x,t,z)e^{i(cx+dt)j} \left[
\begin{array}{c} 1  \\ g_1(z)
\end{array}  \right].
\end{equation}
From (\ref{4.17}) and (\ref{5.9}) it follows that
\begin{equation} \label{5.18}
\frac{d }{d x}S(x,t)^{-1}=-S(x,t)^{-1}\Pi(x,t) \Pi(x,t)^*
S(x,t)^{-1}.
\end{equation}
Hence we obtain
\begin{equation} \label{5.19}
\int_0^{\infty}S(x,t)^{-1}\Pi(x,t) \Pi(x,t)^* S(x,t)^{-1}d x \leq
S(0,t)^{-1},
\end{equation}
i.e., the columns of $\Pi^*S^{-1}$ belong $L^2_2$. As according to
(\ref{5.14}) we have
\[
h(z)-(z-c)=-c^2 \big(z-c+h(z) \big)^{-1},
\]
so taking into account (\ref{5.4}) and (\ref{5.5}) one can see
that
\begin{equation} \label{5.20}
\lim_{x \to \infty}e^{i h(z) x} \Pi(x)=0, \quad \Im \, z > \max
\,(c^2, \, ||Q_1||+1).
\end{equation}
Using the definition of $w_A$, (\ref{5.19}), and (\ref{5.20}) we
derive that the columns in the right-hand side of (\ref{5.17}) are
squarely summable. Thus we have shown that $\wt W(x,t,z)K^*
\O(t,z) \in L^2_2(0, \, \infty)$, and it remains to prove that
$\O_1 \not=0$.

From (\ref{4.21'}) it easily follows \cite{SaL2, SaL3} that
\[
w_A(x,t,z)^*j w_A(x,t,z)=
\]
\begin{equation} \label{5.21}
j-i(z- \ov{z}) \Pi(x,t)^*(A^*- \ov{z}I_n)^{-1}S(x,t)^{-1}(A-z
I_n)^{-1} \Pi(x,t).
\end{equation}
By (\ref{5.9}) and (\ref{5.21}) we have
\begin{equation} \label{5.22}
w_A(x,t,z)^*j w_A(x,t,z) \geq j \quad (z \in \BC_+).
\end{equation}
Notice now that for sufficiently large $\Im z$ the inequality
\begin{equation} \label{5.23}
[1 \quad 0]Z^*e^{-idtj}j e^{idtj} Z \left[
\begin{array}{c} 1 \\ 0
\end{array}  \right]>0
\end{equation}
is true. Thus in view of (\ref{5.12}), (\ref{5.22}), and
(\ref{5.23}) it follows that $\O J \O^*>0$, and so $\O_1 \not=0$.
Therefore equality (\ref{5.11}) is true for all $z$ with
sufficiently large imaginary part and hence for all $z \in \BC_+$.
\end{proof}

\end{document}